# Attacking and Improving the Tor Directory Protocol

Zhongtang Luo
*Purdue University*
luo401@purdue.edu

Adithya Bhat
*Purdue University*
haxolotl.research@gmail.com

Kartik Nayak
*Duke University*
kartik@cs.duke.edu

Aniket Kate
*Purdue University / Supra Research*
aniket@purdue.edu

*Abstract*—The Tor network enhances clients' privacy by routing traffic through an overlay network of volunteered intermediate relays. Tor employs a distributed protocol among nine hard-coded Directory Authority (DA) servers to securely disseminate information about these relays to produce a new consensus document every hour. With a straightforward voting mechanism to ensure consistency, the protocol is expected to be secure even when a minority of those authorities get compromised. However, the current consensus protocol is flawed: it allows an equivocation attack that enables only a single compromised authority to create a valid consensus document with malicious relays. Importantly the vulnerability is not innocuous: We demonstrate that the compromised authority can effectively trick a targeted client into using the equivocated consensus document in an undetectable manner. Moreover, even if we have archived Tor consensus documents available since its beginning, we cannot be sure that no client was ever tricked.

We propose a two-stage solution to deal with this exploit. In the short term, we have developed and deployed TorEq, a monitor to detect such exploits reactively: the Tor clients can refer to the monitor before updating the consensus to ensure no equivocation. To solve the problem proactively, we first define the Tor DA consensus problem as the interactive consistency (IC) problem from the distributed computing literature. We then design DirCast, a novel secure Byzantine Broadcast protocol that requires minimal code change from the current Tor DA code base. Our protocol has near-optimal efficiency that uses optimistically five rounds and at most nine rounds to reach an agreement in the current nine-authority system. Our solutions are practical: our performance analysis shows that our monitor can detect equivocations without changing the authorities' code in five minutes; the secure IC protocol can generate up to 500 consensus documents per hour in a real-world scenario. We are communicating with the Tor security team to incorporate the solutions into the Tor project.

## 1. Introduction

The Tor network is the most widely used anonymous communication solution consisting of over 8,000 volunteered intermediate relays and serves over two million daily users over the past year [30]. Here, a Tor client software on a user device first chooses a small subset of (typically three) relays from all volunteered relays following Tor's path selection strategy [31]. The client then forms a Tor circuit [27] involving cryptographic key exchange and establishing networking sessions over the chosen nodes and routes the user's traffic over this circuit during the session.

The efficiency and security of the Tor service critically depend upon Tor's path selection strategy [5], [13], [35], [36]. There are nine Directory Authorities (DAs) in Tor providing current relay information to the clients. To reduce bandwidth cost so that every client does not have to ask each authority one by one for information, the authorities execute a distributed protocol in which they would all sign on a specific version of the relay parameters/information, called a *consensus document*, to guarantee that the information is as accurate and current as possible, which can then be delivered and verified by any client in one piece.

The consensus documents are vital for Tor to function. Without a way to produce and verify relay information, the client may be tricked into using adversarial relays that easily compromise the anonymity of the client. However, consensus documents differ greatly from the information each authority offers [12], and Tor DAs have occasionally been unable to form a consensus document. In August 2010, the unexpected behavior of one authority triggered a failure in the process of building a consensus document [24]. In January 2021, a sudden distributed DoS attack on authorities caused Tor to be unavailable for several hours due to being unable to form a consensus document [22]. These incidents demonstrate how Byzantine behaviors by even a single authority can render the current Tor system unusable and how a re-examination of the current protocol for generating the consensus documents is necessary to ensure the security of the system.

As our first contribution, we examine the premise of the Tor DA protocol based on source code and documentation and confirm the system model to be standard point-to-point bounded-synchronous communication links with a minority of Byzantine faults (Section 2). We then detail vulnerabilities that can compromise the anonymity of clients. We find that a single compromised authority can create a consensus document not known to correct DA servers by sending different information to different correct authorities. We call this exploit the equivocation attack. Without the client actively checking on all nodes, the exploit remains unnoticed. Furthermore, because of design oversights in the current codebase, we discover practical exploits that compromise the clients' anonymity as an adversary can inject adversarial relay information into this consensus document. To avoid overburdening authorities, clients obtain consensus documents by asking a random authority and trust the document if its signatures are valid. However, an adversary can compromise the client's anonymity by tricking them into using relays that the adversary controls (Section 3).

As our second contribution, we have built TorEq, a temporary remedy service to detect and address the exploit in a reactive manner. After the authorities build a consensus document, the service pulls consensus information from every DA to ensure no equivocation attack occurs. We deployed the service and found out that the service can summarize the information in 5 minutes after the publication of every consensus document (Section 4).

Our third contribution focuses on proactive measures. To ensure that similar exploits do not succeed in the future, we hope not only to fix the specific design oversights that allow the adversary to inject bad relay information but also to create a consensus protocol for Tor that is provably secure in the defined system model. Due to the complex nature of computing one consensus document, we establish that Interactive Consistency [10] is a practical way to implement the Tor DA consensus process. We provide an Interactive Consistency protocol using parallel Byzantine Broadcasts. Since there are only nine DA servers in Tor, we design DirCast, a protocol that improves the efficiency for the low number of servers. Improving the idea from the current state-of-the-art protocol [1], our protocol is highly efficient in the common scenario with no equivocation and terminates in 5 rounds. Our protocol consists of a Bootstrap Phase and an Agreement Phase. In the Bootstrap Phase, each authority broadcasts its information to everyone, and in the Agreement Phase, authorities synchronize with each other to ensure no equivocations occur (Section 5).

We built a test version of the protocol with the current approach in mind, and there is only a small code change (about 8%) from the original code base. We evaluate our protocol in geographically distributed environments like Tor and show they are highly applicable: our protocol can generate up to 500 consensus documents per hour, and Tor only needs one consensus document per hour. Our protocol is also transparent to clients, so only an update for the Directory Authorities is necessary to switch to the new protocol (Section 6).

**Responsible Disclosure.** We contacted the Tor security team on this vulnerability on April 27, 2022. The security team acknowledged the issue on May 6, 2022. We had further communications during the summer of 2022, including an in-person meeting. We simultaneously developed an equivocation detector as an emergency measure (TorEq). It was merged into Tor's codebase on August 11, 2023 [38].

As the Tor team is working towards migrating Tor from C to Rust in Project Arti [34] and still developing support for onion services as of 2023, other features such as Directory Authorities are currently not under active development, but we are looking forward to working with the team to implement our protocol offering proactive security.

## 2. Preliminaries

We begin by introducing the Tor Directory service, its goals, and the assumption. We then present the current protocol that we obtained by analyzing the source code.

### 2.1. Background of the Directory Protocol

Since Tor's service depends on relays run by volunteers, one essential aspect of utilizing such a service is that the client has a list of relays so that the client can choose three to establish the circuit. Serving a correct and current list of relays to every client is a security matter directly related to the guarantee of anonymity: the client's anonymity depends on the fact that the relays do not track the client collaboratively. An adversary that controls the first relay can see who is accessing the network and start fingerprinting; an adversary that controls the third relay can see what is being accessed. Therefore, we must ensure the client gets correct and current information on relays.

Naïve solutions are not ideal in this scenario: hard coding every relay into the software is impractical, as relays run by volunteers may not remain usable throughout the update cycle of the Tor software. Another naïve solution is to have the Tor client fetch a list of the information of relays from an authoritative source, like Tor's website, before it tries to establish any circuit. This solution establishes a single point of failure: if the website is down, then Tor may not be usable. Any compromise to the authoritative source also directly impacts Tor's anonymity, while any hacker that controls the source can compromise Tor.

To address the concern, Tor set up a distributed system consisting of 9 special relays as Directory Authorities (DA) to help collect and disseminate information on relays. While the client technically can ask each of the authorities for a list of relays and aggregate them locally, at the current size of the list (approximately 2 MB each, according to estimation on the statistic data [30]), each update would require approximately 20 MB of data. With the number of clients utilizing the system (around 2 million at any given time in 2021 [30]), it is more bandwidth-efficient if the client only needs to fetch one consensus document on aggregated relay information instead of 9 separate documents. This reason drives the Tor team to develop a Directory Protocol that allows the authorities to produce and agree on one consensus document every hour for the clients to use.

The protocol used by Tor DA has been reworked several times. The current Version 3 Directory Protocol has a detailed specification that outlines the expected behavior of an authority [29]. Tor is also an open-source project and publishes its code including the Directory Authority portion for examination [32].

As of 2021, the 9 Directory Authorities publish one consensus document every whole hour. Every consensus document is valid for 3 hours, which is designed to provide redundancy. Relays use HTTP POST to send their descriptor to the authorities. Both relays and clients fetch the current consensus document via HTTP GET.

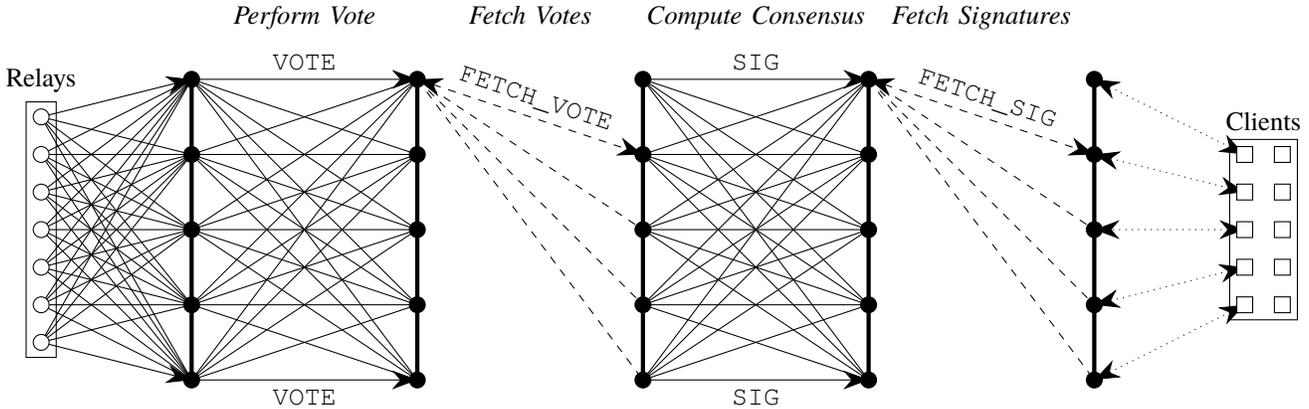

Figure 1: Current Tor's voting scheme. Thick lines represent messages that will be sent. Dashed lines represent potential communications if the first authority cannot hear from some authority. Dotted lines represent communications with the client. Outline of the protocol: (1) Every relay sends its information to the authorities when it boots up. (2) In the *Perform Vote* step, every authority shares what it knows to every other authority in the form of a vote. (3) In the *Fetch Votes* step, if one authority has not heard from someone, it asks everyone else if they have heard from it. (4) In the *Compute Consensus* step, every authority builds consensus locally, signs it and broadcasts the signature. (5) In the *Fetch Signatures* step, if one authority has not received the signature from someone, it asks everyone else for it. (6) When it is time to publish a consensus document, Authorities collect the signatures and then generate the consensus.

## 2.2. System Model and Goals

Within the specification of Tor [29], authorities are expected to publish one consensus document every hour. Furthermore, the current Tor protocol divides the ten-minute consensus document generation process into four 150-second rounds, in which each authority is expected to act according to the specification. We conclude that the specification resembles a bounded synchrony assumption [3]. Furthermore, as the specification states that every authority ensures accurate clocks, we assume the presence of a global clock, whose skew is largely irrelevant because the protocol is run sparsely.

According to the specification, the current Tor protocol tries to fix an issue from the previous versions: to 'prevent authorities from lying' [29]. Therefore, the threat model allows incorrect authorities controlled by the adversary to behave arbitrarily.

Summarizing from above, we consider the system model of Tor to be a standard bounded synchronous and authenticated setting. In this setting, the adversary may take control of some authorities less than half. This setting is common in works and applications dealing with consensus [1], [3], [6], [23].

The protocol should satisfy safety and liveness requirements when facing an adversary controlling a minor number of authorities. From a broad viewpoint, safety is the clients' anonymity not being compromised due to the consensus document, and liveness is the clients' timely receiving of the consensus document. While an accurate formal description that captures these ideas is difficult to obtain, we claim that the following properties should hold even if a minor number of authorities behave arbitrarily:

(1) **Agreement.** All correct authorities should output the same consensus document determined by some aggregation rule.
(2) **Integrity.** The adversary should not be able to forge a valid consensus document different from what the correct authorities output.
(3) **Liveness/Termination.** The protocol should produce one consensus document each hour.

## 2.3. Current Protocol

Since the current description of the Directory Protocol does not specify distributed system behavior [29], we analyze the source code [32] of Tor and summarize the behavior of authorities below.

There are 4 rounds in the current consensus protocol: Perform Vote, Fetch Votes, Compute Consensus, and Fetch Signatures. Each round lasts 150 seconds. Therefore, authorities start running the protocol 10 minutes before the publication of a consensus document. Figure 1 displays a visual representation of the protocol.

Below we provide a detailed description of the protocol. For clarity, we consider a system with authorities $\{P_1, P_2, \ldots, P_n\}$ and relays $\{r_1, r_2, \ldots, r_m\}$.

(1) **Perform vote.** At the start of each consensus protocol, each authority $P_i$ collects some meta-data (voting interval, length of each step, etc.) and information on every relay $r_x$ it knows about. It then packs the information of the relays into a vote $V_i$ and broadcasts the vote VOTE($V_i$) to every other authority.
(2) **Fetch missing votes.** If an authority $P_i$ has not received a vote from authority $P_k$ in the previous round, it broadcasts a request FETCH_VOTE($i, k$) to every other authority. Upon receiving any such request, the authority sends back the corresponding vote to the authority if it has received one from $k$ in the previous round. After receiving the vote, the authority checks the vote and keeps the latest vote in record according to the timestamp on the vote.

(3) **Compute consensus.** Each authority builds a consensus document locally based on the votes it received. Formally, given a list of votes $L_i = (V_1, V_2, ..., V_n)$, the authority $P_i$ summarizes information on every relay $r_x$ to build a consensus document of available relays $C_i = (r_1, r_2, ..., r_m)$ based on the following procedure:
  (a) The authority checks if at least $\lfloor \frac{n}{2} \rfloor + 1$ votes have been received. It aborts if there are insufficient votes.
  (b) For each relay $r_x \in \bigcup_{V \in L_i} V$:
    (i) The authority first determines if it should be in the list. A relay $r_x$ is in the list the authority publishes if and only if it is included in at least $\lfloor \frac{n}{2} \rfloor + 1$ votes.
    (ii) If there is a naming conflict, the authority takes notice and picks the name from the authority with the largest ID.
    (iii) For each property of $r_x$, determine the value from the most popular opinion, i.e., the one that is favored by most votes.
      (A) This includes all the flags (Exit, Guard, Running, BadExit, MiddleOnly, Valid). In case of a tie, the flag is not set.
      (B) This also includes the version and protocol of the relay. In case of a tie, the largest version and/or protocol is selected.
      (C) The authority determines bandwidth by the median of all votes that contain them. It only acknowledges measured bandwidth if there are no less than 3 authorities that measured it.
      (D) In case of a tie, it uses the lexicographically larger one as the exit policy summary.

It then signs the consensus document and broadcasts the signature $\sigma(C_i)$.

(4) **Fetch missing signatures.** If an authority $P_i$ has not received a signature from authority $P_k$ in the previous round, it broadcasts a request `FETCH_SIG`$(i,k)$ to every other authority. Upon receiving any such request, the authority sends back the corresponding signature to the authority if it has received one from $P_k$ in the previous round. After receiving the signature the authority $P_i$ checks the signature to see if it matches $C_i$ and adds it to the consensus document.

Once it is time to publish a consensus document, the authority checks if $\lfloor \frac{n}{2} \rfloor + 1$ signatures have been collected. If so, the authority makes the consensus available to the public.

Tor Proposal 207 [17] outlines the concept of directory guards as a means for the clients to fetch consensus documents but is not specific on the implementation. We examine the official source code [32] and conclude that in the current implementation, every Tor relay fetches the consensus document from a random authority via HTTP every hour. Every Tor client instead fetches the consensus document from a random source based on a hard-coded list of relays and authorities via HTTP at boot time. It then keeps using the document until it is no longer fresh (after one hour), after which the client will repeat the fetching process from one of its entry guards [33] for a new document.

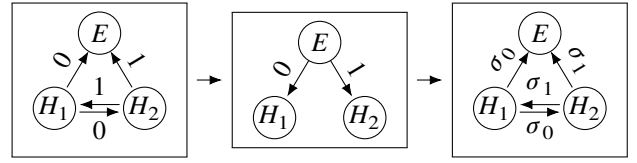

Figure 2: Example of an equivocation attack. The adversary party $E$ can cheat by sending different values to $H_1$ on the left and $H_2$ on the right and trick them into signing on different values. 1) $E$ receives 0 from $H_1$ and 1 from $H_2$ as they broadcast their respective input. 2) $E$ equivocates by sending 0 to $H_1$ and 1 to $H_2$. 3) $H_1$ sees that both $E$ and $H_1$ have input 0 and thus signs on 0 as the protocol output. $H_2$ sees that both $E$ and $H_2$ have input 1 and thus signs on 1 as the protocol output. $H_1$ will think that $H_2$ has misbehaved and disregard the 'incorrect' signatures from $H_2$. $H_2$ will also think that $H_1$ has misbehaved.

## 3. Vulnerabilities of the System

Intuitively the vulnerability of the current protocol is that it does not address equivocations of adversarial authorities. We analyze how the adversary can utilize the vulnerability to evade the current consensus monitoring mechanism and poison any client based on the official implementation. We then study three possible attack vectors that allow us to compromise the liveness or integrity of the protocol with a minority (even one) adversarial authority.

### 3.1. Equivocation in the Current Protocol

An equivocation attack happens when authorities send different messages to different correct authorities. Since the current protocol lacks any equivocation check, it is easy for adversarial authorities to violate the safety property and create two valid consensus documents.

Consider a simple scenario: the 9 Tor Directory Authorities $P = \{P_1, P_2, ..., P_9\}$ need to agree on whether a particular relay should be listed in the consensus document. Each of the 9 authorities $P_i$ has a value $x_i$ of either 0 or 1, where 0 means that the authority believes that the relay should not be listed, and 1 means that the authority considers that the relay should be listed in the consensus document.

Recall that the authorities need to sign the same consensus document, so they need to agree on one value of $x$. In the current protocol, each authority collects every $x_i$ from every authority $P_i$ via a vote in the Perform Vote step and signs the value that appears more often. This allows an equivocation exploit. Assume a minority adversary party of 3 authorities $E = \{P_1, P_2, P_3\}$ knows the values of honest authorities $x_4 = x_5 = x_6 = 0$ and $x_7 = x_8 = x_9 = 1$. They can tell correct authorities $H_1 = \{P_4, P_5, P_6\}$ that $x_1 = x_2 = x_3 = 0$ by voting, so $H_1$ sees 6 values of 0 and signs $x = 0$ in the Compute Consensus step. They can then tell correct authorities $H_2 = \{P_7, P_8, P_9\}$ that $x_1 = x_2 = x_3 = 1$ by voting, so $H_2$ sees 6 values of 1 and signs $x = 1$. The process is demonstrated in Figure 2.

Furthermore, the adversary can now create two versions of the consensus at the same time, as they now have 3 signatures on both $x = 0$ and $x = 1$. They can sign both $x = 0$ and $x = 1$ to get another 3 signatures for both values, so they now have 6 signatures - a majority - on both $x = 0$ and $x = 1$, and create two contradictory consensus documents based on the signatures. It is now possible for the adversary to broadcast only the signatures for $x = 0$ in the Compute Consensus step, so the public only sees the consensus document with $x = 0$ and 6 signatures. The consensus document with $x = 1$ has only 3 signatures (less than half of the authorities), so it will not be published by the authorities that signed on $x = 1$. However, the adversary has the consensus document with $x = 1$ and 6 signatures, so it can feed a client it wants to poison with this document. The client with this document will see $x = 1$ and acts differently than other clients that see $x = 0$.

This exploit is possible for even one adversarial authority as long as the initial values of correct authorities are evenly split between $x = 0$ and $x = 1$.

## 3.2. Utilizing the Exploit

Because the current Tor consensus health monitor [28] collects only the consensus document and the votes from the respective voters, the monitor cannot detect the exploit as long as adversarial authorities also provide one version of the vote to the monitor. Furthermore, even if we have all the archived Tor consensus documents available, we do not know if such an attack happened in the past, because the attack is not observable on the correct consensus document except for missing signatures.

For a liveness attack, we can apply the equivocation attack directly since our goal is to disrupt the generation of any consensus document. For an integrity attack, we aim to feed the client with an incorrect consensus document so that the client uses the document and loses anonymity.

The client, by default, picks a source from a hard-coded list uniformly at random to ask for the current consensus document at boot time. If the source is adversarial, that source can provide the client with an incorrect consensus document. Meanwhile, if the source is not adversarial, since Tor uses HTTP for all transmissions, an adversary can also spoof connections to authorities and provide the client with the incorrect consensus document. As the client does not store archived consensus, the exploit will also not be traceable by the client once it fetches a new consensus and dumps the current one.

For this exploit to be practical, the adversary needs some capability to manipulate the input of the correct authorities to achieve meaningful results. We detail three possible attack vectors below.

```
/* Pick a bandwidth */
if (num_mbws > 2) {
  rs_out.has_bandwidth = 1;
  rs_out.bw_is_unmeasured = 0;
  rs_out.bandwidth_kb =
    median_uint32(measured_bws_kb, num_mbws);
} else if (num_bandwidths > 0) {
  rs_out.has_bandwidth = 1;
  rs_out.bw_is_unmeasured = 1;
  rs_out.bandwidth_kb =
    median_uint32(bandwidths_kb, num_bandwidths);
  if (n_authorities_measuring_bandwidth > 2) {
    /* Cap non-measured bandwidths. */
    if (rs_out.bandwidth_kb >
        max_unmeasured_bw_kb) {
      rs_out.bandwidth_kb = max_unmeasured_bw_kb;
    }
  }
}
```

Figure 3: Code of bandwidth computation. Line 2 shows that the bandwidth is considered measured if 3 authorities have measured it.

## 3.3. Attack Vectors

**3.3.1. Liveness Attack.** The current computation process of the consensus document is sensitive to input from one party in several aspects: one party with the largest ID can determine the name of any relay, and because the bandwidth of any relay is determined by the median value of all votes, one adversarial party can also equivocate on the bandwidth to cause different correct parties to compute different consensus documents.

This sensitivity is sufficient to halt the execution of Tor with the application of an equivocation attack. One adversarial authority can manipulate the scheme such that every correct authority compute a different consensus document from each other. In this scenario, no consensus can be made since every authority has a different output.

**3.3.2. Sybil Relay Injection.** A Sybil attack [37] is a common exploit to deanonymize Tor's clients. Since the client's anonymity is broken if all three relays the client uses work to track it collaboratively, in this exploit, an adversarial party tries to register as many relays as possible to gain a disproportionately large influence and trick clients into using them.

Because Sybil attacks against Tor are tracked by manual data analysis instead of automated process [20], an equivocation attack helps greatly by preventing the consensus document with Sybil relays from being seen by the public.

To use the equivocation attack, one adversarial authority can keep the Sybil relays unknown to some correct authorities but known to others to cause a consensus split. This is done by having the Sybil relays broadcast to only part of the correct authorities.

**3.3.3. Bandwidth Manipulation.** In the current protocol, the consensus acknowledges measured bandwidth as long as 3 authorities measure it. (See Figure 3.)

Meanwhile, if an adversarial relay is newly introduced to the system, it will initially have no measured bandwidth since nobody has measured it. Therefore, 3 adversarial authorities can work together to publish a bandwidth before real measurements are taken, and the incorrect bandwidth will be included in the final consensus document.

In the current Tor system, the bandwidth of a relay is proportional to the probability of the client using it. Therefore, if one adversarial relay has incorrect but high bandwidth, the client will use it with high probability.

### 3.4. Attack Demonstration

For demonstration purposes, we pick a situation where a party of 3 adversarial authorities adds relays with arbitrary bandwidth and creates two consensus documents.

**3.4.1. Steps of the Exploit.** We denote the party of 3 adversarial authorities $E$. We also arbitrarily decide a party of $r = \lfloor \frac{n}{2} \rfloor - 2$ correct authorities to be manipulated to sign our incorrect consensus document beforehand so that we reach a quorum of $f+r = \lfloor \frac{n}{2} \rfloor + 1$. Denote this party of correct authorities $H'$. Denote the rest of the correct authorities $H$.
(1) **Prepare the list of relays.** Denote $C$ to be what we would normally vote during the Perform Vote step and $C'$ to be the vote with the modified measured bandwidth.
(2) **Actions during the Perform Vote step.** Authorities in $E$ would vote $C'$ to the party $H'$ and $C$ to the party $H$.
(3) **Actions during the Compute Consensus step.** Authorities in $H'$ would sign on $C'$ and authorities in $H$ would sign on $C$. As long as authorities in $E$ publish their signatures on $C$, it would be the consensus document for the round.
(4) **Create an unpublished consensus.** Authorities in $E$ can sign on $C'$ privately. Together with the signatures from $H'$ this would result in an unpublished consensus on $C'$.

**3.4.2. Experimentation.** We used Chutney [26] to simulate a Tor network with 9 authorities, similar to the current layout of Tor. With a modified codebase for the adversarial authorities, we show that 2 correct authorities in party $H'$ signed on the incorrect consensus document but could not publish it due to a lack of signature, while other authorities agreed on the correct version of the consensus document. (See Figure 4.)

## 4. Reactive Measure

In this section, we offer a reactive measure that detects the equivocation scenario. The proposed measure is easy to integrate into the current Tor network and offers fast and reliable detection, though it does not prevent the exploit altogether.

### 4.1. Introduction of TorEq

While the Tor development team is working on developing the next iteration of the software in Rust, the current

```
r test010r kNeiqbQSrPh/JPuJiTrcz1bNDTY Nf2VyvkI...
2022-04-05 17:27:05 127.0.0.1 5010 0
......
w Bandwidth=14597871
......
-----BEGIN SIGNATURE-----
KtR7wLvxNtat1Kly71bjJVyWp9gwuPbggnQYBdZI8dWLm7M...
......
-----END SIGNATURE-----
```

```
Apr 05 13:27:20.657 [warn] A consensus needs 5 good
signatures from recognized authorities for us to
accept it. This ns one has 2 (test003a test004a).
7 (test005a test000a test006a test002a test007a
test008a test001a) of the authorities we know
didn't sign it.
```

Figure 4: Demonstration of the bandwidth exploit. This is an example of a correct authority that signs on the incorrect consensus but cannot publish it. The top box displays the consensus that the authority signed. Note the unusually high bandwidth. The bottom box displays the log of the authority.

Figure 5: A typical display of the monitor plugin. The image above is a display of a real-world scenario where no equivocation happens; the image below is a test-case scenario where an equivocation happens.

system still needs to address the issue. Therefore, we have developed TorEq, a plugin to be put into the Tor Consensus Health Monitor using the current Python codebase of the monitor. The monitor generates one web page per hour to track the consensus document and the consensus protocol. The plugin we develop displays the vote each authority receives from other authorities and automatically checks for discrepancies. Figure 5 shows a typical display of the plugin.

The plugin automatically collects every vote that every authority receives concurrently every hour, which has a communication overhead of $n^2d$, where $n$ is the number of authorities (9 as of now) and $d$ is the document length. It then compares them to see if any authority has equivocated and provides a warning if it finds such behavior.

While this is a short-term fix for the system, it does not entirely prevent the exploit, and equivocation is still possible.

Introducing a single monitor places trust in the monitor, which becomes a single point of failure and may also suffer DDOS attacks that render it unavailable. Any adversary that can corrupt the authorities and the monitor can bypass the detection altogether. Meanwhile, the reactive measure means that we cannot provide a remedy automatically if we detect an equivocation, and the Tor service may not be safe until we resolve the situation manually.

In case of a detected equivocation, we recommend the clients continue using the latest consensus document known to be safe (generated without equivocating). While the client may cache such consensus document over continued usage, we recognize there might be no feasible way to obtain the latest safe consensus document during bootstrapping for a new client. Therefore, we recommend that the authorities keep track of the equivocation situation and always keep the latest safe consensus document available.

### 4.2. Benchmark

For our plugin to be efficient in a practical setting, we need to demonstrate that it can generate a result in a short time in comparison to the 3-hour lifetime of any consensus document. It should also be usable under the setting of a typical server machine that the Tor consensus monitor uses.

**Benchmark Setup.** We tested our consensus monitor with our plugin on a typical server setting: a t3a.medium AWS instance with 4 GB RAM, 8 GB hard disk, 2 vCPUs running at up to 3.3 GHz, and a bandwidth of up to 5 Gbps. We developed our plugin based on the open-sourced monitor codebase in Python [28], and we use the current monitor as a baseline to compare the performance. We run both monitors' web page generation process 5 times to output a web page displaying the consensus document and consensus protocol status, fetching data from the real-life Directory Authorities. We compared our monitor's performance with the current insecure version that does not detect equivocation.

**Benchmark Result.** We discover that our monitor consumes 140.7 MB bandwidth, compared to 18.0 MB of the current monitor. This is to be expected, as our monitor has a higher communication complexity than the original monitor of $n^2 d$ ($n$ is the number of authorities and $d$ is the length of the document). Our monitor generates a web page in 229.164 seconds on average, while the current monitor generates one in 134.988 seconds. As the process needs to be done once an hour, we conclude that our solution runs in a comparable time to the current insecure monitor and is practical.

## 5. Improving the Protocol

Next, we examine what a consensus protocol like the one Tor is using wants to achieve and design a secure protocol that satisfies Tor's needs.

### 5.1. Definition

The current aggregation rule needs to compute different subjects, including the presence of a relay, its flags, its bandwidth, etc. (see Section 2.3) To implement the rule, one reasonable way is to aim for Interactive Consistency, which allows an authority to compute the consensus document locally in the same way as the current design:

**Definition 5.1** (Interactive Consistency (IC) [8]). *With a system of n servers $\{P_1, P_2, \ldots, P_n\}$ and each server $P_i$ starting with a value $x_i$, the following properties hold after the protocol execution:*
*(1) **Termination.** Every correct server $P_i$ eventually outputs a vector $X_i = \{x_{i,1}, x_{i,2}, \ldots, x_{i,n}\}$ of size n.*
*(2) **Agreement.** If server $P_i$ and $P_j$ are correct, then $X_i = X_j$.*
*(3) **Validity.** If a correct server $P_i$ starts with the value $x_i$, then $x_{i,i} = x_i$.*

Tor's current consensus protocol resembles an attempt to achieve IC. However, in the current protocol, adversarial authorities can compromise the protocol by equivocating during the broadcast. Hence, the essential vulnerability of the system is the insecure broadcast channel, which allows the adversary to equivocate and create an incorrect consensus document that can compromise the anonymity of the client. Therefore, it is important to improve the protocol to achieve a secure broadcast, which is defined as the Byzantine Broadcast problem.

**Definition 5.2** (Byzantine Broadcast (BB) [4]). *With a system of n servers $P = \{P_1, P_2, \ldots, P_n\}$, and the sender $P_s \in P$ starting with a value x, the following properties hold:*
*(1) **Termination.** Each correct server $P_i$ eventually outputs a value $x_i$.*
*(2) **Agreement.** If server $P_i$ and $P_j$ are correct, then $x_i = x_j$.*
*(3) **Validity.** If the sender $P_s$ is correct, then every correct server outputs x.*

We can design a protocol that achieves IC with $f$ adversarial servers out of an $n$-server system as long as $n \geq f + 1$, by utilizing $n$ authenticated broadcast channels [21]. However, considering that the authorities output a consensus document signed by the majority eventually, and an adversarial majority can fabricate any document with the signatures of a majority, we limit our discussion to situations where $n \geq 2f + 1$. A secure BB protocol solves the current vulnerabilities Tor has. Therefore, we suggest a new version of the BB protocol for the current consensus protocol to use.

### 5.2. Considerations on the Tor System

While there are a few secure BB protocols, such as the state-of-the-art protocol [1], the specific scenario of Tor motivates us to design a new protocol that achieves the best performance under the situation.

*1) A very low number of nodes.* As of now, Tor has 9 Directory Authorities. The number is unlikely to increase significantly even in the coming years, given that the most recent admission is in 2017 [31]. Most BB protocols focus on the common scenario where hundreds of nodes need to talk to each other and emphasize performance in such a scenario. For example, we notice that the state-of-the-art protocol requires on average 10 rounds to achieve consensus regardless of the number of nodes - that is more rounds than the number of nodes in the Tor system.

*2) An emphasis on having fewer rounds.* The current system of Tor assumes significant network delay, as the current round time is set to 2.5 minutes. The new protocol needs to achieve a low number of rounds, especially in the most common scenario where all nodes are correct, without compromising its ability to achieve consensus when the adversary controls some of the nodes.

*3) A lack of cryptographic infrastructure.* Many BB protocols, such as the state-of-the-art protocol [1], assume cryptographic infrastructure such as a common-coin scheme. We observe that Tor does not have such a built-in cryptographic infrastructure, and implementing one with some common method [6] takes at least 5 rounds of communication, which makes it a liability on performance.

### 5.3. Key Ideas

We observe that designing a good protocol to solve the problem is not always easy: a naïve solution would be to have every authority forward every vote it receives to every other authority. However, the solution does not prevent the adversarial authorities from equivocating in the forwarding and causing a consensus split.

Fortunately, we do not have to design a BB protocol from scratch. $O(1)$ round complexity can be achieved with the state-of-the-art protocol designed by Abraham et al. [1] The state-of-the-art protocol utilizes the idea of a core iteration to achieve $O(1)$ round complexity. We summarize the execution of the core iteration with $n$ servers and $f$ Byzantine faults such that $n \geq 2f + 1$ below:
(1) **Propose Round.** The sender signs and broadcasts its value.
(2) **Vote Round.** Every server signs and broadcasts the value it has received from the sender to everyone else as a vote.
   At the end of the round, each server checks the vote it receives. It commits to a value $x$ if it receives at least $f + 1$ votes on $x$ and receives no other properly signed values, i.e. detects no equivocation. It then records all signatures on the vote on $x$ as a certificate $C$.
(3) **Notify Round.** Every server, that commits to a value, signs and broadcasts a notification message with $x$ and $C$.
(4) **Termination.** Once a server receives $f + 1$ notification messages on the same value $x$. It broadcasts the signatures of $f + 1$ notify messages and terminates on $x$.

The core iteration algorithm above has many ideal properties for building a BB protocol: (1) All correct servers terminate on the correct value if the sender is correct in an iteration. (2) If one correct server terminates, all correct servers terminate in the next round. (3) No correct servers terminate on different values. As we assume that there are few Byzantine actors, termination in 4 rounds is ideal. The issue with the core iteration is that the correct servers might not terminate after the iteration finishes. However, we can integrate a less efficient phase that only uses the committed value and its certificate to ensure termination.

### 5.4. DirCast: Our Core Protocol

We designed DirCast, an authenticated synchronous BB protocol specifically tuned for the Directory Protocol in Tor, assuming at most $f$ adversarial servers out of $n \geq 2f + 1$ servers. The protocol employs ideas from [1], [9] that allow us to achieve high efficiency when the sender is correct.

DirCast consists of two phases: the Bootstrap Phase and the Agreement Phase. During the Bootstrap Phase, the sender proposes its value to every server. They also exchange the messages they have received among them. Upon finishing the Bootstrap Phase, at most one unique value is committed with a valid certificate, and each server can use the certificate to prove the validity of the committed value. During the Agreement Phase, each server synchronizes the intention among them to make sure everyone stays consistent.

We describe our protocol below. Figure 6 gives a visual representation of DirCast.

**Description of the Protocol.** Denote the set of servers participating in the protocol $P = \{P_1, P_2, \ldots, P_n\}$. Denote the sender is $P_s \in P$. Every server $P_i \in P$ prepares an empty vector $L_i$ at the start of the protocol to record the vote it has received.

**Bootstrap Phase.** In the Bootstrap Phase, the sender broadcasts the message, and every other server relays the message to ensure delivery.
(1) **Propose Round.** The sender $P_s$ signs the value $x$ to get the signature $\sigma_s(x)$ and broadcasts PROPOSE$(x, \sigma_s(x))$ to every server, including itself.
(2) **Vote Round.** At the start of the Vote Round, if a server $P_i \in P$ has received message PROPOSE$(x, \sigma_s(x))$ with the correct signature in the Propose Round, it signs the value to get the signature $\sigma_i(x)$ and broadcasts message VOTE$(x, \sigma_s(x), \sigma_i(x))$ to every server. In case $P_i$ receives multiple PROPOSE messages of different values, it sends the first two values as two VOTE messages separately. If a server $P_i \in P$ receives from another server $P_j \in P$ its vote message VOTE$(x, \sigma_s(x), \sigma_j(x))$ for the first time, server $P_i$ adds $(x, \sigma_s(x), \sigma_j(x))$ to its vote vector $L_i$.

At the end of the Vote Round, every server $P_i \in P$ sets its commit state $x_i \leftarrow (x, \sigma_s(x))$ if and only if:
1) $L_i$ consists of only $x$ and its signatures.
2) $|L_i| \geq f + 1$.
Otherwise, server $P_i$ sets $x_i \leftarrow (\bot, \bot)$. In other words, the server $P_i$ commits if and only if it has received at least $f + 1$ votes and detects no equivocation. If a server commits, it also stores the signatures on the $f + 1$ votes as a certificate $C(x)$.

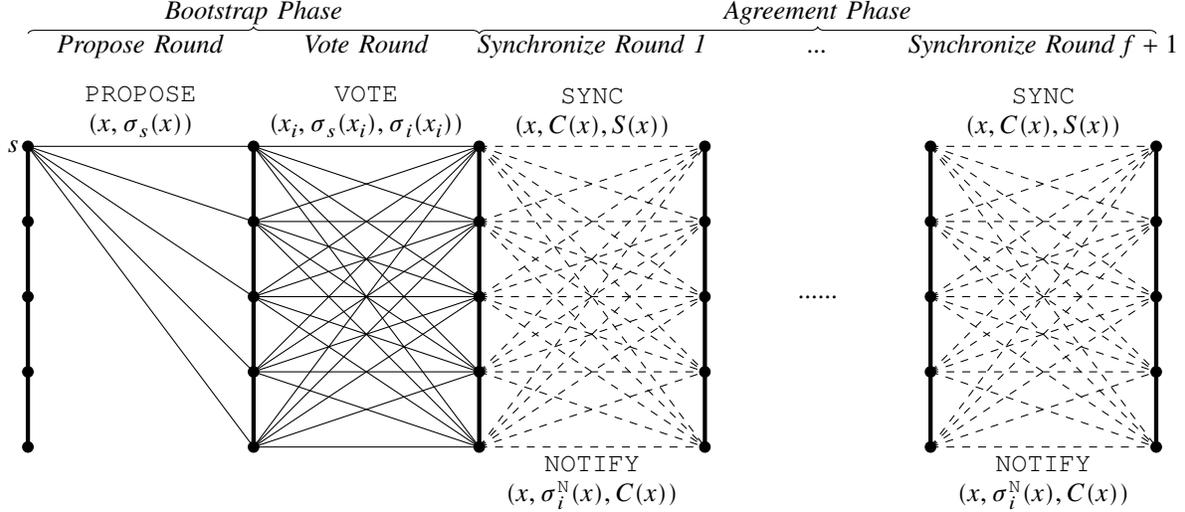

Figure 6: Visualization of the core protocol. Thick lines represent messages that will be sent. Dashed lines represent potential communications. Outline of the protocol: (1) In the *Propose Round*, the sender sends its value to everyone. (2) In the *Vote Round*, everyone sends the value it has received to everyone. (3) Throughout the *Synchronize Rounds*, SYNC messages and NOTIFY messages are propagated to ensure all servers agree on the same value. If the sender is correct, all correct servers terminate after 2 rounds, else it can take up to $f + 1$ rounds for a server to terminate.

**Agreement Phase.** In the Agreement Phase, each server $P_i \in P$ records all certificates $(x, C(x))$ and all signatures on the certificates $S(x) = \{\sigma(C(x))\}$ it has seen. At the start of the phase, $P_i$ only sees a certificate if it has committed to a non-empty value, and there are no signatures on any certificates. Then, we run the Synchronize Round $f + 1$ times:

(1) **Synchronize Round.** At the start of the $t \in [1, f+1]$-th Synchronize Round, each server $P_i \in P$ that has sent less than two different SYNC messages checks for values and certificates it received that have exactly $t - 1$ signatures in $S(x)$ and have not been signed by it. If there are such certificates, it signs them so that there are $t$ signatures and broadcasts $\text{SYNC}(x, C(x), S(x))$ to every other server. If a server $P_i \in P$ receives from a server $P_j \in P$ a message $\text{SYNC}(x, C(x), S(x))$ with $x$ not known to $P_i$, server $P_i$ checks that $|S(x)| = t$. If that is the case, server $P_i$ marks the value and the certificate as received.

(2) **Early termination.** At the start of the Agreement Phase, every server $P_i \in P$ that commits to a non-empty value $x$ signs the NOTIFY message and broadcasts $\text{NOTIFY}(x, \{\sigma_i^N(x)\}, C(x))$.

Once a server has received $f + 1$ NOTIFY messages on the same value $x$, it broadcasts the signatures of these NOTIFY messages and terminates on $x$. This ensures that all the other servers receive these signatures and terminate on the next turn.

At the end of $f + 1$ Synchronize Rounds, every server $P_i \in P$ that has not terminated via early termination terminates with $x$ if it has seen one and only one value $x$. Otherwise, it terminates with $\bot$.

The pseudocode of the protocol is available below:

**State variables:**
$n \leftarrow \text{GetN}()$ ▷ Number of authorities
$f \leftarrow \lfloor \frac{n-1}{2} \rfloor$ ▷ Number of adversarial authorities
$id \leftarrow \text{GetId}()$ ▷ Index of the server
$s \leftarrow \text{GetSender}()$ ▷ Index of the sender
$\Delta \leftarrow \text{GetRoundTime}()$ ▷ Round time
$livePropose \leftarrow []$ ▷ All Propose messages
$voteValues \leftarrow \text{Map}()$ ▷ All Vote values and sigs
$commitValue \leftarrow \bot$ ▷ Committed value
$commitCert \leftarrow \{\}$ ▷ Certificate of the value
$notifyValues \leftarrow \text{Map}()$ ▷ All Notify values
$syncValues \leftarrow \text{Map}()$ ▷ All SYNC values
$syncMsgSent \leftarrow 0$ ▷ How many SYNC is sent
$finalValues \leftarrow \{\}$ ▷ All values received

**procedure** GetRound($t$)
  **if** $t \leq \Delta$ **then return** *Propose*
  **else if** $t \leq 2\Delta$ **then return** *Vote*
  **else if** $t \leq (f+3)\Delta$ **then return** $Sync(\lfloor \frac{t}{\Delta} - 1 \rfloor)$
  **return** $\bot$

**procedure** Propose
  $x \leftarrow \text{GetInitialValue}()$
  $\sigma_s(x) \leftarrow \text{Sign}(x)$
  Broadcast $\text{PROPOSE}(x, \sigma_s(x))$

**upon** receiving a valid $\text{PROPOSE}(x, \sigma_s(x))$ **do**
  **if** GetRound($currentTime$) = *Propose* **then**
    $livePropose \leftarrow livePropose \cup \{(x, \sigma_s(x))\}$

**procedure** Vote
  $u \leftarrow \text{Min}(|livePropose|, 2)$
  **for** the first $u$ elements $(x, \sigma_s(x)) \in livePropose$ **do**
    $\sigma_{id}(x) \leftarrow \text{Sign}(x)$
    Broadcast $\text{Vote}(x, \sigma_s(x), \sigma_{id}(x))$

**upon** receiving a valid $\text{VOTE}(x, \sigma_s(x), \sigma_i(x))$ **do**
  **if** GetRound($currentTime$) = *Vote* **then**
    $voteValues[x] \leftarrow voteValues[x] \cup \{\sigma_i(x)\}$

**procedure** Commit
  **if** $|voteValues| = 1$ **then**
    $x \leftarrow$ the only value in $voteValues$
    **if** $|voteValues[x]| \geq f + 1$ **then**
      $commitValue \leftarrow x$

```
        commitCert ← {voteValues[x]}
procedure NOTIFY
    if commitValue ≠ ⊥ then
        x ← commitValue, C(x) ← commitCert
        σ_{id}^N(x) ← SIGN(NOTIFY(x))
        Broadcast NOTIFY(x, {σ_{id}^N(x)}, C(x))
upon receiving a valid NOTIFY(x, {σ^N(x)}, C(x)) do
    notifyValues[x] ← notifyValues[x] ∪ {σ^N(x)}
    if |notifyValues[x]| ≥ f + 1 then
        Broadcast NOTIFY(x, notifyValues[x], C(x))
        Output x and terminate
procedure INITSYNC
    if commitValue ≠ ⊥ then
        x ← commitValue, C(x) ← commitCert
        σ_{id}(C(x)) ← SIGN(C(x))
        syncValues[x, C(x)] ← {}
        finalValues ← {x}
procedure SYNC(r)                    ▷ r-th Synchronize Round
    for (x, C(x)) ∈ syncValues do
        if syncMsgSent ≤ 2 and |syncValues[x, C(x)]| = r − 1
then
            σ_{id}(C(x)) ← SIGN(C(x))
            syncValues[x, C(x)]  ←  syncValues[x, C(x)] ∪
{σ_{id}(C(x))}
            Broadcast SYNC(x, C(x), syncValues[x, C(x)])
            Remove (x, C(x)) from syncValues
            syncMsgSent ← syncMsgSent + 1
upon receiving a valid SYNC(x, C(x), S(x)) do
    if GETROUND(currentTime) = Sync(|S(x)|) then
        syncValues[x, C(x)] ← S(x)
        finalValues ← finalValues ∪ {x}
upon GETROUND(currentTime) = Propose do
    if id = s then
        PROPOSE()
upon GETROUND(currentTime) = Vote do
    VOTE()
upon GETROUND(currentTime) = Sync(r) do
    if r = 1 then
        COMMIT()
        NOTIFY()
        INITSYNC()
    SYNC(r)
upon GETROUND(currentTime) =⊥ do
    if |finalValues| = 1 then
        x ← the only value of finalValues
        Output x and terminate
    Output ⊥ and terminate
```

## 5.5. Security Proof

We prove that DirCast satisfies the properties of a BB protocol: Termination, Validity, and Consistency.

### 5.5.1. Termination. We show that DirCast terminates within a certain time frame.

**Theorem 5.1.** *DirCast terminates in at most $f + 3$ rounds.*

*Proof.* It is trivial to observe that the protocol terminates after 1 Propose Round, 1 Vote Round and at most $f + 1$ Synchronize Rounds. Therefore, it terminates in at most $f + 3$ rounds. □

### 5.5.2. Validity. We show that DirCast correctly delivers the message when the sender is correct.

**Theorem 5.2.** *If the sender $P_s$ is correct and sends $x$, every correct server $P_i$ terminates with $x$ via early termination.*

*Proof.* Since the sender is correct, every correct server receives the same Propose message with $x$ at the end of the Propose Round. Therefore, there will be at least $f + 1$ votes for $x$ during the Vote Round and no equivocation. By definition, every correct server broadcasts a NOTIFY message at the start of the Agreement Phase, and they all terminate early with $x$. □

### 5.5.3. Agreement. We show that DirCast is secure against equivocation attacks, starting by showing that the Agreement Phase is consistent within itself with a Dolev-Strong style reasoning:

**Lemma 5.1.** *If a correct server $P_i$ receives a valid SYNC message with value $x$, $x$ is broadcast to every server in a valid SYNC message at the end of the protocol.*

*Proof.* If $P_i$ received SYNC($x$) before round $f + 1$, $P_i$ will broadcast the message.

If $P_i$ received SYNC($x$) during round $f + 1$, since there are $f + 1$ signatures on the message, one correct server $P_j$ must have received $x$ before round $f + 1$, and thus $P_j$ have broadcast the message. □

**Lemma 5.2.** *If a correct server $P_i$ terminates non-early on $x_i$ and a correct server $P_j$ terminates non-early on $x_j$, then $x_i = x_j$.*

*Proof.* From the previous lemma, we can conclude that all correct servers received the same set of SYNC messages. Therefore, $P_i$ and $P_j$ should follow the same termination rule. Hence, $x_i = x_j$. □

We then show that any termination is consistent.

**Lemma 5.3.** *If some correct server $P_i$ terminates early on $x \neq \perp$, then there exists some correct server $P_j$ that has committed on $x$ at the end of the Bootstrap Phase.*

*Proof.* Since $P_i$ received $f + 1$ NOTIFY messages, one message must have come from a correct server $P_j$. That server has committed on $x$ at the end of the Bootstrap Phase. □

**Lemma 5.4.** *If some correct server $P_i$ commits on $x \neq \perp$, then no correct server has voted for some value $x' \neq x$.*

*Proof.* Since $P_i$ commits on $x$, it received no votes on any value other than $x$. Therefore, no correct node has broadcast a vote on a value other than $x$. □

**Lemma 5.5.** *If some correct server $P_i$ commits on $x \neq \perp$, then every correct server terminates on $x$.*

*Proof.* Since $P_i$ commits on $x$, by Lemma 5.4 no correct server has voted for some value $x' \neq x$. Hence no correct server has received a SYNC or NOTIFY message with some value $x' \neq x$ as for such message to be valid, $C(x')$ needs $f + 1$ votes. Therefore, if a correct server terminates early, it terminates on $x$.

Since by definition of the protocol $P_i$ broadcasts a SYNC message with $x$, every correct node receives this message. Therefore, every correct server terminates non-early terminates on $x$. □

**Theorem 5.3.** *If a correct server $P_i$ terminates with $x_i$ and a correct server $P_j$ terminates with $x_j$, then $x_i = x_j$.*

*Proof.* If neither $P_i$ nor $P_j$ terminates early, then by Lemma 5.2 we know $x_i = x_j$.

If at least one of the servers, say $P_i$, terminates early on $x_i$, then by Lemma 5.3 some server committed on $x_i$. Therefore, by Lemma 5.5 $x_i = x_j$. □

### 5.6. Adapting DirCast to the System

Since the Perform Vote and the Fetch Vote rounds in the original protocol resemble $n$ BB protocols, if we run $n$ core protocols concurrently in place of these steps, we can achieve a version of the protocol with safety guarantees without modifying how each authority aggregates the consensus document.

**Equivocation Detection.** Our design allows authorities to output if they detect an equivocation happening within the system. They can output the two conflicting votes in the same hour as evidence that allows investigation into the cause.

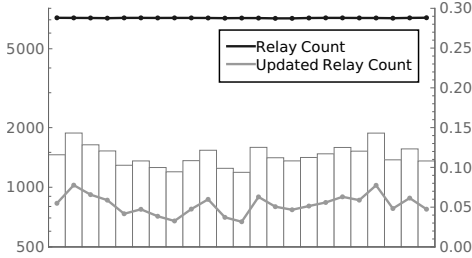

Figure 7: Number of relays and updates on June 1, 2022. The lines represent the number of relays on the logarithmic scale and the bars represent the percentage of relays updated since the last consensus.

**Message Compression.** To make up for the increased traffic due to the protocol, we propose one way for authorities to compress their message when communicating with each other. From the consensus document data we collected, we discovered that there is little difference between the authorities' vote and the consensus from the last hour - less than 15 percent of the relays have different information. Figure 7 shows the number of relays on a typical day. Therefore, during communication between authorities, it is more efficient for the authorities to only transmit information on relays that have updated information from the last consensus document. This compression method is safe against equivocation because the last consensus document is unique and available.

We also notice that only the digest of the vote needs to be transferred during the agreement phase instead of the full document, since any valid SYNC and NOTIFY message must have a value that has been seen by every server in the Bootstrap Phase. This discovery allows us to further reduce communication complexity in practice.

TABLE 1: The table displays the round type, its length (i.e., how many rounds of this type during a broadcast) (Co.), its communication overhead in terms of signature length and document length (Comm.), and the cryptographic overhead (number of sign operations) over one broadcast (Crypto.). $d$ - document length, in proportion to the number of relays. $\kappa$ - signature length, currently 512 bytes after base-64 encoding. $f$ - number of adversarial authorities the protocol can tolerate. $n$ - number of authorities. **eq.** - equivocation, i.e. the sender is adversarial. **sim.** - simultaneously, i.e. the Synchronize Rounds and the Early Termination run concurrently, so if one terminates the other will also terminate. **E.T.** - Early Termination.

| Type | Co. | Comm. | Crypto. |
|---|---|---|---|
| Propose | 1 | $(d + \kappa)n$ | 1 |
| Vote | 1 | $(d + 2\kappa)n^2$ | $n$ |
| Sync (w/o eq.) | sim. | $[d + (f+2)\kappa]n^2$ | $n$ |
| E.T. (w/o eq.) | 2 | $[d + (2f+3)\kappa]n^2$ | $n$ |
| Sync (w/ eq.) | $f$ | $[2d + 2(f+1)\kappa]n^2$ | $2n$ |
| E.T. (w/ eq.) | sim. | $[d + (f+2)\kappa]n^2$ | 0 |

TABLE 2: Communication overhead of the protocol. The table displays the protocol (Broadcast - one broadcast. C.D. - one consensus document generation with $n$ broadcasts), the round complexity (Ro.), the communication overhead in terms of signature length and document length (Comm.), and the cryptographic overhead (number of sign operations) (Crypto.). Assuming $t$ equivocations in Consensus (w/ eq.). Notations are the same as in the previous table.

| Protocol | Ro. | Comm. | Crypto. |
|---|---|---|---|
| Broadcast (w/o eq.) | 4 | $(3n^2 + n)d +$ $[(3f+7)n^2 + n]\kappa$ | $3n + 1$ |
| C.D. (w/o eq.) | 5 | $(3n^3 + n^2)d +$ $[(3f+7)n^3 + 2n^2]\kappa$ | $3n^2 + n$ |
| Broadcast (w/ eq.) | $f + 3$ | $(4n^2 + n)d +$ $[(3f+6)n^2 + n]\kappa$ | $3n + 1$ |
| C.D. (w/ eq.) | $f + 4$ | $(3n^3 + n^2t + n^2)d +$ $[(3f+7)n^3 + (2-t)n^2]\kappa$ | $3n^2 + n$ |

### 5.7. Analysis of DirCast

We summarize the round complexity and the communication overhead in Table 1 and Table 2.

**Round Complexity.** For the core protocol, if the sender is correct, the protocol terminates in 4 rounds, better than previous $(f + 1)$-round simple protocols [9]. If the sender is adversarial then the protocol terminates in at most $f + 3$ rounds, which is 8 for a 9-server system. This is better than the current state-of-the-art system designed to have a constant number of rounds (currently 10) for a larger set of servers [1].

After DirCast, one more round is needed for each authority to collect all the signatures and publish the consensus document. Therefore, after adapting our core protocol, the whole consensus protocol terminates in 5 rounds in the best-case scenario and $f + 4$ rounds in the worst-case scenario.

**Communication Overhead.** To generate a consensus document for $n$ authorities with up to $f$ faulty authorities, our protocol has a communication complexity of $O(n^3(d+f\kappa))$ with or without equivocation from the sender. Here, a document size is $d$ and a signature size is $\kappa$. Further methods, such as extension protocols [16], can be developed on top of the protocol to reduce communication complexity at the cost of protocol complexity.

## 6. Benchmark of DirCast

There are many questions we want to answer for our protocol: (1) Is our protocol practical in the real world? Can it be used by Tor authorities and generate consensus documents on time? (2) Is our protocol efficient? How much network and CPU time does it need to generate a consensus? (3) Is our protocol scalable? How do we fare if the number of relays continues to increase? To answer these questions, we test our protocol compared to the current one based on latency, throughput, and scalability.

### 6.1. Benchmark Setup

We implemented a prototype version of our protocol on the existing Tor C code base. Due to many similarities, the implementation is light-weight with about 1,000 lines of code addition out of 13,000 lines of code in the Directory Authority module, about 8 percent, as we took over the Fetch Vote round in the original protocol and repurposed it to the Agreement Phase. The prototype code is available at https://github.com/zhtluo/DirCast.

### 6.2. Protocol Parameters

We set the number of authorities to reflect the current situation of Tor, i.e. 9 authorities. We assumed that a quorum is correct according to the protocol requirement and set the protocol to tolerate up to 4 faults in the system.

In the real-world scenario, Tor authorities run one round in 2.5 minutes and finish the protocol with 4 rounds in 10 minutes. They start the protocol and generate the consensus document once every hour. We adjusted the round time for the experiment to be shorter while ensuring that all communications could still be completed within the time frame. Based on the experiment we set the time length of one round to be 24 seconds, and run the protocol once every 4 minutes.

We varied the number of relays up to 3,000 in the system during the experiment, as the number of relays directly affects the size of each message and hence impacts the performance. We boot all these relays on an outside server to simulate the message exchange with the authorities. We also assumed that 15 percent of the relays updates their information in every consensus document.

### 6.3. Server Specifications

We set up our experiment in a way that resembles real-world Tor Directory Authority distribution. As of 2022, Tor

TABLE 3: Geographic Location of the Current Tor Directory Authorities

| Name | Location |
| --- | --- |
| dizum | Netherlands |
| dannenberg | Germany |
| tor26 | Austria |
| bastet | United States |
| maatuska | Sweden |
| moria1 | United States |
| Faravahar | United States |
| longclaw | Canada |

Directory Authorities have 5 nodes in Europe and 4 nodes in North America (see Table 3). Therefore, we used 9 AWS instances, of which 5 are in *eu-central-1* location and 4 are in *us-east-2* location.

We assumed that the authorities had limited computing powers and network bandwidth, as in the real world they would also have to handle connections with clients. Therefore, we used t2.micro AWS instances with 1 GB RAM, 8 GB hard disk, 1 vCPU running at up to 3.3 GHz, and a bandwidth of 60-80 MBits/s for all authorities.

### 6.4. Test Cases

We tested our protocol's performance in three different situations with the test network. We first tested the current insecure protocol on the network with the above-suggested parameters and specifications. This served as a baseline for measuring our protocol with the current behavior of the Tor network. We then tested our secure protocol with all authorities behaving correctly. This is the expected scenario in the real world. We also tested the scenario where one authority tried to split the consensus by equivocating, which helped to demonstrate the security of the protocol and allowed us to benchmark the situation when an exploit happens. We consider the scenario where multiple authorities equivocate to be similar to the one-equivocation situation, as the adversary has no significant advantage in this scenario over just equivocating once.

### 6.5. Benchmark Criteria

**6.5.1. Latency.** Latency refers to the time between the start of the protocol and the generation of the document. For a lock-step setting such as Tor, latency depends on the round time and the number of rounds. The round time is somewhat arbitrary both in our benchmark and in the real-world scenario as long as it covers the network and CPU time, but it still depends on how robust we want the protocol to be against fluctuating network delay. However, by keeping the round length consistent throughout our benchmark process, we gauge how our protocol performs concerning the current insecure protocol assuming a similar network environment.

**Methodology.** We fixed the round time in the test scenario to 24 seconds. We booted 3,000 relays in the system and let 15% of the relays change their information every time the protocol starts. We bootstrapped authorities to run the protocol in test repeatedly until all relays are in the consensus document.

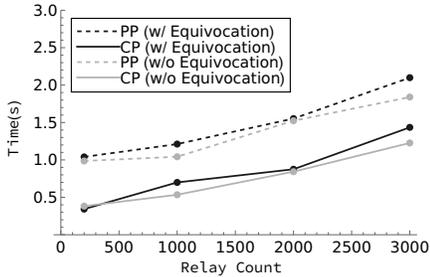 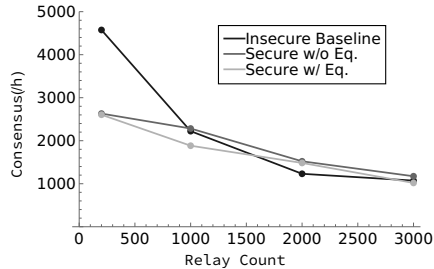 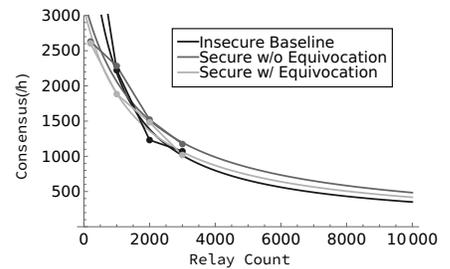

Figure 8: Average network and CPU time for different phases of our protocol. PP - Bootstrap Phase. CP - Agreement Phase.

Figure 9: Network throughput in three test scenarios, measured in consensus per hour.

Figure 10: Prediction of throughput up to 10,000 relays. The protocol can generate up to 500 consensus documents per hour.

**Result and Analysis.** During our experiment, our protocol was able to generate a consensus document with a 24-second round time. Due to the similar round complexity, our protocol finished in time comparable to the original insecure protocol when all authorities behave correctly and there is no equivocation, with a 25 percent increase in latency. Our protocol finished in 120 seconds and the original protocol finished in 96 seconds. Meanwhile, even if an authority equivocates, we are still able to finish our protocol in 216 seconds with a relay count of 3,000 in the test scenario.

In the real-world scenario, if we use the current Tor configuration of 2.5 minute per round, our protocol finishes in 12.5 minutes when there is no equivocation, and 22.5 minutes with equivocation. This is within the one-hour mark, so our protocol finishes on time if we use the 2.5-minute round time.

**6.5.2. Throughput.** Throughput refers to how many consensus documents we can generate in a certain amount of time, assuming that we run as many test protocols as we want in parallel. While we do not need to run multiple protocols in parallel in the real world, this test helps us to answer how efficient our protocol is, i.e. how much system resource it takes to produce one consensus document.

**Methodology.** We varied the number of relays between 200, 1,000, 2,000, and 3,000. We let 15% of the relays change their information every time the protocol starts. We measured each phase's network and CPU time in the protocol and computed the throughput from the measurements.

**Result and Analysis.** We plotted the average CPU and network time of each phase in detail in Figure 8. We observe that the Agreement Phase takes roughly one-third of the network and CPU time of the whole protocol. While equivocation by one authority increases the message count of the protocol, each authority sends at most 2 `SYNC` messages in the Agreement Phase in one broadcast, so equivocation does not impact throughput significantly. Figure 9 shows the network's throughput in the three test cases. We discovered that equivocation by one authority does not significantly decrease throughput, although it does increase latency.

We conclude that our protocol runs in comparable network and CPU time to the current protocol because we compressed our votes based on the consensus last round (see Section 5.6), drastically cutting message size.

**6.5.3. Scalability.** For scalability, we consider it unlikely that Tor will get new authorities in the foreseeable future. Therefore, our main concern is the ever-increasing number of relays in the system, and we predicted how the protocols will fare as the number of relays increases. This helps us visualize how our protocol will work.

**Methodology.** From the analysis of our protocol, the communication overhead scales linearly with the size of the document. Therefore, we perform a linear regression on each protocol's network and CPU time. We then compute the throughput based on the regression result, up to 10,000 relays.

**Result and Analysis.** We observe that with the increase in the number of relays, the round complexity of our protocol stays the same, but the communication complexity scales linearly as the document size increases. This observation allows us to plot and predict the performance of our protocol up to 10,000 relays, a 25 percent increase in the amount of the real-world situation now.

From Figure 10, we conclude that our protocol can generate up to 500 consensuses per hour with a relay count of 10,000, comparable to the current insecure protocol. We conclude that our protocol is practical to use in real-world scenarios.

## 7. Related Work

### 7.1. Improvements on the Tor Consensus Protocol

There have been a few Tor proposals for making the Tor consensus protocol more robust. Some proposals were implemented, while some are in discussion. However, to the best of our understanding, no proposal addresses the equivocation problem and the resulting vulnerabilities discovered in this paper.

Since Proposal 206 [19] was implemented, every Tor client fetches the consensus document from a hard-coded list of relays and authorities at boot time. This proposal is designed to alleviate the bandwidth overload of authorities but does not address the underlying security issue: an adversarial server can still equivocate and serve clients and/or relays with an incorrect consensus document.

Proposal 207 [17] allows the client to use directory guards (which are the same as entry guards in the current implementation) in the client's current consensus document to fetch an updated consensus document. The motivation of the proposal is to reduce fingerprinting of client IPs done by relays and authorities that serve consensus documents. The proposal does not address the equivocation problem: an adversary is still able to create equivocating documents, and clients at boot time, as well as relays, are still vulnerable to being served with an incorrect consensus document.

Proposal 239 [18] proposes that every consensus document includes a list of hashes of previous consensus documents. While the proposal is still open for discussion, it does not fully solve the equivocation vulnerability: the incorrect consensus document can still include hashes of previous correct consensus documents, and the adversary can create a forked chain of consensus documents to avoid detection. While the client may find out it has been compromised later via a correct consensus document on some other fork, the damage is already done.

Proposal 267 [15] proposes to establish consensus transparency like certificate transparency [14] that allows consensus to be traced to ensure validity. However, much of the proposal is left unclear (there are 26 instances of TODO/TBD in the original proposal, including both the security discussion and the implementation specifications). In the proposal, authorities submit their consensus to a public logger to be logged. We think that depending on the exact implementation, the logger can also become a central point of failure. Ultimately, to ensure that the consensus is consistent in a distributed system, we cannot sidestep the Byzantine Agreement problem, the main problem we discuss in this paper.

### 7.2. Byzantine Agreement (BA)

To generate Tor's consensus document is to solve the Byzantine Agreement (BA) problem on all relay parameters. The problem informally requires that (i) all correct servers with inputs output the same value, and (ii) if the inputs of correct servers are all $x$, then the output of the correct servers is $x$.

Abraham et al. [1] showed a randomized protocol in a synchronous and authenticated setting, which achieves expected $O(1)$ round complexity and $O(n^2)$ communication complexity, and $O(f)$ round complexity with a deterministic protocol [1]. These are the state-of-the-art approaches for BA.

There are several variants of BA. One variant uses strong validity, in which the output must be a value proposed by at least one of the correct servers [11]. Another variant uses median validity, in which the output of the protocol must span the inputs of the correct servers [25]. The decision about the inclusion of a relay uses everyone's inputs, and the computation of the bandwidths uses medians (see Section 2.3). Hence, these works give another technique to generate Tor consensus documents.

However, the authorities must agree on many values in the consensus document: a list of relays, each relay's flags, version, protocol, bandwidth, and exit policy. Only some values, such as a list of relays and bandwidth, can be merged using such protocols. Therefore, we decided not to design a protocol that deals with these values individually but to design a protocol that achieves Interactive Consistency and lets each authority compute the consensus document locally instead.

### 7.3. Interactive Consistency (IC) and Byzantine Broadcast (BB)

As we have reasoned in Section 5, the current consensus protocol is an attempt to achieve Interactive Consistency (IC) (Definition 5.1). Achieving IC in a $n$-server system through $n$ simultaneous Byzantine Broadcasts (BBs) (Definition 5.2) is a common approach [10] that we have decided to adopt.

The Dolev-Strong protocol [9] is a classic BB protocol with $f + 1$ round complexity. The paper also showed that deterministic protocols tolerating $f$ faults must have at least $f+1$ rounds. Our protocol is better than the traditional Dolev-Strong protocol regarding optimistic latency. The advantage comes with little overhead: even though our protocol is more costly than previous protocols such as Dolev-Strong [9] in terms of using more signatures, signature length in the protocol is relatively insignificant compared to document length, so our protocol does not invoke a significant communication complexity increase; meanwhile, our protocol can achieve an optimistic latency decrease from 750 seconds to 600 seconds with the current Tor setting, a 20% improvement. While the Dolev-Strong protocol fares better in the pessimistic case, we consider a failure in the Tor consensus protocol unlikely, so we did not optimize for the situation. To illustrate the point, we compute the total communication complexity in both broadcast protocols, assuming a typical 1000-relay update (see Table 4). We conclude that our protocol incurs a communication overhead of less than 2%, with notable benefit in optimistic latency. Our protocol also supports more authorities without increasing optimistic latency, while Dolev-Strong incurs a linear overhead increase.

The randomized protocol by Abraham et al. [2] is state-of-the-art, achieving optimal early stopping, optimal resilience, and polynomial complexity [2]. Our solution cannot use randomized protocols due to their large cryptographic overheads or the expensive distributed setup requirement.

## 8. Concluding Remarks and Future Work

This paper demonstrates a vulnerability in the deployed consensus mechanism used by Tor DAs. The exploit shows the inherent difficulty in designing a secure consensus protocol for the production environment: even what looks like a simple mechanism can fail in multiple surprising ways and may result in exploits with real-world implications. Furthermore, the exploit can be carried out by one Directory Authority and is not detectable with the current monitor or client, and not observable even if we have archived every consensus document since the beginning of the protocol. To address the issue, we provided TorEq, a short-term remedy that uses a consensus monitor to detect such exploits.

TABLE 4: Comparison between our broadcast protocol and the Dolev-Strong protocol based on a typical 1000-relay update, 9 authorities and a 150-second round time. We use a signature length of 502 bytes, a digest length of 53, and a relay entry length of 337, mirroring real-life Tor consensus documents.

| Overhead | Our Protocol | Dolev-Strong |
| --- | --- | --- |
| Latency (Optimistic) (s) | 600 | 750 |
| Latency (Pessimistic) (s) | 1050 | 750 |
| Communication (Optimistic) (MB) | 31.0 | 30.4 |
| Communication (Pessimistic) (MB) | 31.1 | 30.6 |

We also observe that it is important to design protocols in which safety can be proven. Towards that end, we define the Tor consensus process as the interactive consistency (IC) primitive in the distributed computing world. We then design DirCast, a secure Byzantine broadcast (BB) protocol that could replace the current flawed protocol in use and provides proven security of IC realization based on the BB protocol. The protocol finishes in five rounds in the optimistic scenario and nine rounds in the worst-case scenario in a 9-node DA system. Based on benchmarks, we predict that the protocol can generate up to 500 consensus documents per hour in a geographically distributed system resembling Tor's DAs and a relay size similar to real-world situations. We take special care to ensure that our protocol resembles the current Tor DA consensus protocol as we develop the prototype within 1,000 lines of code. We, therefore, conclude that our design is simple and practical to be utilized in real-world scenarios.

Keeping immediate real-world usability in mind, our proposed solution follows the adversary and network assumption of the current Tor consensus protocol: tolerating any minority of corruptions in the bounded-synchronous setting. Given that Tor needs to run this consensus process once every hour, the bounded-synchrony assumption seems acceptable. Nevertheless, in the future, it can be interesting to design a secure Tor DA consensus protocol in a partial synchronous or asynchronous communication model. However, it will be a drastic departure from state of the art: the number of tolerable failures reduces to a maximum of $n \geq 3f+1$; we cannot solve the IC problem and have to resort to a weaker version called vector consensus [7] where inputs from $f$ honest parties may not be considered. Nevertheless, to deal with the network asynchrony correctly, it can be interesting to define a novel network model and propose the Tor DA consensus process in that model.

## Acknowledgement

This work was supported in part by NIFA award number 2021-67021-34252 and by the National Science Foundation (NSF) under grant CNS1846316. We thank the Tor Team, especially Roger Dingledine and Micah Anderson, for their helpful input and insight. We thank the reviewers of IEEE S&P for their feedback. We would also like to thank Jianting Zhang for his effort in proofreading the paper.

# Appendix A.
# Meta-Review

## A.1. Summary

The Tor anonymous communication system relies on 9 Directory Authorities (DAs) to construct a consensus document containing a list of the network's relays and their properties. This paper demonstrates that the Directory Authority consensus protocol does not guarantee consensus in the malicious setting—a minority set of authorities can launch an equivocation attack that may produce two distinct, valid consensus documents. Tor clients rely on this consensus document for critical functions, such as choosing overlay network paths, and the authors explore how client security may be compromised as a result of equivocation. This paper constructs two possible solutions to the attack: (1) TorEq, an external consensus monitoring system; and (2) DirCast, a byzantine broadcast protocol that the authorities can run in parallel to provably reach consensus.

## A.2. Scientific Contributions

- Creates a New Tool to Enable Future Science
- Identifies an Impactful Vulnerability
- Provides a Valuable Step Forward in an Established Field

## A.3. Reasons for Acceptance

1) This paper clearly identifies a crucial flaw in Tor's consensus mechanism and demonstrates the vulnerability with a safe, simulated experiment using real code. Tor is an important privacy tool with millions of daily users, so the finding is significant. The implications of the flaw are explained clearly and convincingly.
2) To our knowledge, this work is the first to (1) view the Tor consensus mechanism as a formal distributed systems problem, (2) suggest notions of security, and (3) apply a Byzantine Broadcast protocol to reach consensus.
3) The tools and code developed for this paper are available for use and will likely make Tor safer in the near term. We hope this paper motivates Tor's adoption of a provably correct consensus protocol such as DirCast.